# Electric field and photo-excited control of the carrier concentration in GdN

H. Warring, B.J. Ruck, H.J. Trodahl and F. Natali

*The MacDiarmid Institute for Advanced Materials and Nanotechnology, Victoria University of Wellington, PO Box 600, Wellington 6140, New Zealand*

**Abstract**

We present both electric-field and photo-excited control of the carrier concentration in GdN. There is no evidence in the results of a carrier-mediated contribution to the Gd-Gd exchange interaction that has been suggested to explain a measured Curie temperature that is much higher than obtained within theoretical treatments. Persistent carrier concentrations seen in both the field-effect and photo-induced conductivities point to a distribution of long-lived trap states below the conduction band, very likely centred at nitrogen vacancies.

The rare-earth nitrides (RENs) have recently been re-examined as intrinsic ferromagnetic semiconductors, a class of material that has relatively few members. Theoretical advances now permit detailed predictions concerning their transport and magnetic states, though without to date a great number of experimental results to support them.[1-3] However there is a slowly advancing experimental literature based largely on the growth of high-quality thin films.[4-14] Within the series GdN, the most thoroughly studied, is the member for which the 4*f* shell is exactly half filled, with a net spin of 7 $\mu_B$ and zero net orbital moment. A Curie temperature ($T_C$) near 70 K establishes its RE-RE exchange as the strongest within the series.[4,5,15] It has been suggested that a homogeneous ferromagnetic (FM) phase exists only below 50 K, and that the commonly quoted 70 K $T_C$ signals a FM phase nucleated by magnetic polarons near nitrogen vacancies,[16] but even that 50 K is higher than the $T_C$ for any other member of the REN series. The promise of GdN has been demonstrated recently in a superconductor-GdN-superconductor tunnel junction that provides excellent spin selectivity.[17] Furthermore GdN islands have been shown to enhance the efficiency of GaN tunnel junctions.[18]

To gain a full understanding of the material, especially in regard to its action in devices, it is essential to investigate the effects of doping control of the carrier density. The RENs are all susceptible to the formation of nitrogen vacancies ($V_N$), which dope them *n*-type.[5,19] To date the lowest carrier concentration in epitaxial films is of order $10^{20}$ cm$^{-3}$, apparently limited by the equilibrium concentration of $V_N$ at the temperatures required for epitaxial growth.[20] Films grown at lower temperature have been reported with reduced carrier densities, but at the expense of a polycrystalline structure.[4,5] No attempts at compensation by co-doping with acceptors have been so far reported. Nonetheless there is an urgency to explore the effects of carrier concentration, both for the further development of possible devices and to determine whether free-carrier-mediated exchange might resolve a disagreement between the predicted < 20 K $T_C$ and the measured 50-70 K.[21,22] There are indeed reports of enhanced conductivities and Curie temperatures at very high $V_N$ concentrations, of order and larger than 10%,[23,24] but at such high $V_N$ concentrations there is the potential that the effects are related to structural rather than purely doping differences. Thus in the present work we report an investigation in which the carrier density is influenced both by a field effect and by photoexcitation.

The films used here were grown in an ultra-high vacuum system at ambient temperature by depositing Gd at 0.05 nm s$^{-1}$ in an atmosphere of $10^{-4}$ Torr of pure nitrogen, and capped with a GaN layer to passivate them against reaction with air. These growth conditions lead to resistivities of around 0.5 Ω cm at room temperature in strongly (111) textured polycrystalline films of the bulk GdN lattice constant (0.499 nm) within the 0.2% uncertainty of the XRD measurement. The XRD linewidths yield crystallite diameters of typically 12 nm. Films used for photoconductivity studies were grown on sapphire substrates and Ni(3nm)/Au(50nm) or Cr(3nm)/Ag(50nm) contacts were made in the **v**an der Pauw geometry by thermal deposition. For field effect studies films were grown on conducting ($\rho$ = 0.018 Ωcm) Si substrates that served as the gate electrode, with a 300 nm insulating layer of SiO$_2$ as the dielectric. The Cr(3nm)/Ag(50nm) contacts for this experiment were deposited before the GdN layer, with the final structure then as shown in Figure 1(a).

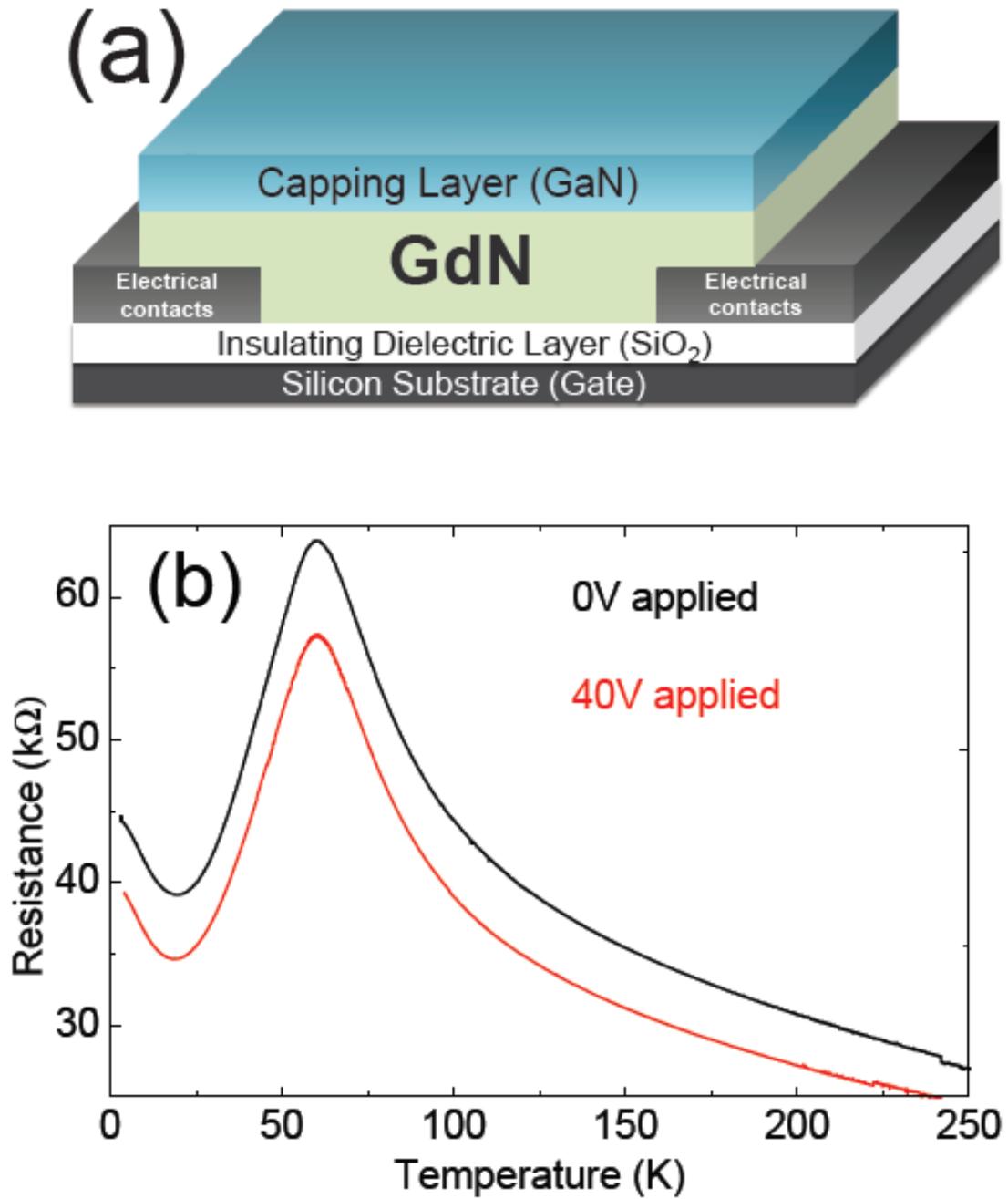

**Figure 1**: (a) Schematic diagram of the field-effect transistor structure used to vary the carrier concentration of the GdN layer. (b) Temperature dependent resistance of a field-effect transistor measured with gate biases of 0 V and +40 V (accumulation). A 40 V bias across the 300 nm $SiO_2$ dielectric corresponds to an induced carrier increase of $2\times10^{17}$ $cm^{-3}$.

The field-effect control of resistivity shown in Figure 1(b) demonstrates that under a +40 V (accumulation) gate potential the carrier concentration was raised by ~10% across the entire temperature range from ambient temperature to 4 K. The 40 V applied across the 300 nm thick $SiO_2$ dielectric ($\varepsilon = 4$) corresponds to a sheet-charge increase of $3\times10^{11}$ cm$^{-2}$, and a volume concentration increase of $2\times10^{17}$ cm$^{-3}$ in the 150 nm thick channel of this device. The implication is that the as-grown channel had a concentration of $2\times10^{18}$ cm$^{-3}$, in line with our previous observation for GdN layers with resistivities in this range.[5] The resistive anomaly across the Curie temperature is reported in many previous films, with the resistance maximum typically falling below but close to $T_C$.[4,5,23] In these data that maximum lies at 60 K in both the as-grown and accumulated-carrier states, as usual very close to the paramagnetic (i.e., Curie-Weiss law) transition temperature determined from magnetisation measurements on the as-grown film (not shown). There is no evidence at all of an enhanced Curie temperature in the accumulation state that would signal a contribution from carrier-mediated exchange.

Significantly, when the gate was earthed to the source the resistance did not return to the as-grown value until a period of hours had elapsed. There are clearly electron traps close below the conduction-band (CB) edge, providing a continuous source of electrons that reside in the CB between re-trapping events. The electron traps also limit the field-effect transistor (FET) function in depletion mode. Thus to explore the influence of traps more fully, and to investigate the effects of stronger carrier-concentration control, we have performed a set of similar temperature-dependent resistivity studies while enhancing the carrier concentration by photo-excitation.

The band gap in GdN is about 0.5 eV, but it is an indirect-gap semiconductor with the valence band maximum at Γ and the CB minimum at X.[25] The optical gap is determined by direct transitions from a trough in the valence band at the X point to the CB minimum. The absorption onset is 1.3 eV at ambient temperature, falling to 0.9 eV in the spin-split band structure below $T_C$,[8,25-27] so that a HeNe laser serves to activate the conductivity. The absorption coefficient of GdN at 633 nm is approximately $7\times10^4$ cm$^{-1}$,[25] giving an absorption length comparable to the thickness (~150 nm) of the films used here. For these van der Pauw measurements the ~2 mW of 633 nm light incident on sample was diffused to illuminate the entire ~ 5 mm × 5 mm conducting patch.

Figure 2(a) shows the resistivity of a typical film from ambient temperature to 5 K. The solid line corresponds to the as-grown film, and the dashed line was measured under illumination. Directly below ambient temperature the resistivity rises as carriers freeze out of the conduction band with an activation energy of about 20 meV. The data show a peak at the ferromagnetic phase transition of 65 K, followed below 20 K by a continued rise from a shallower donor level. The dashed curve showing the illuminated conductivity exhibits a clear photoconductive response at all temperatures, with an ambient-temperature activation energy reduced to ~15 meV. The increase in conductivity ($\delta\sigma = \sigma - \sigma_0$ where $\sigma$ and $\sigma_0$ are the illuminated and dark conductivity respectively), which is proportional to the number of extra charge carriers excited into the conduction band, is shown in Figure 2(b).

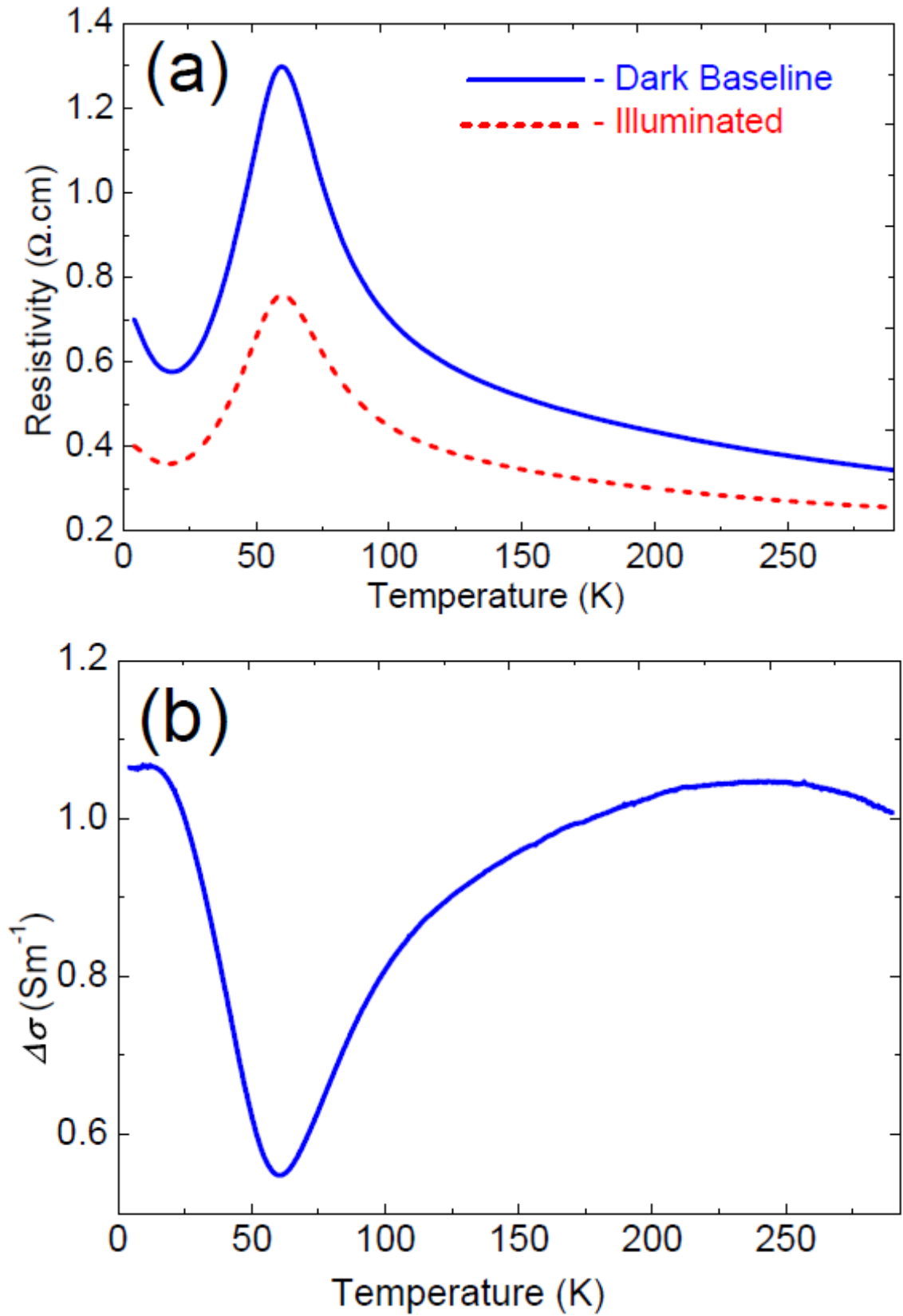

Figure 2: (a) Dark (solid line) and illuminated (dashed line) resistivity of a 150 nm thick GdN film illuminated by 2 mW of 633 nm light. (b) The change in conductivity induced by illumination for the sample shown in (a).

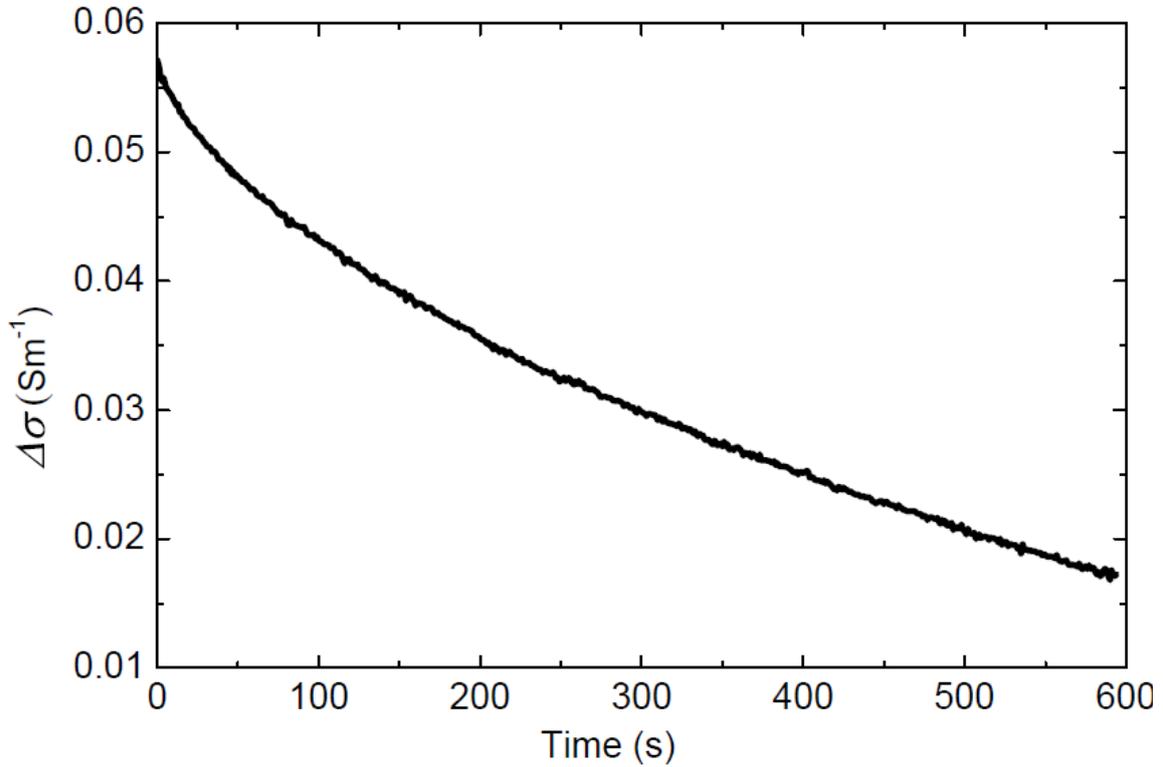

Figure 3: Persistent photoconductivity at room temperature (logarithmic scale). The conductance falls towards the dark value after the illumination is switched off, with a time scale of about 10 minutes. The deviation from exponential recovery is largest at the highest carrier concentration. The recovery is slowed by several orders of magnitude at 10 K.

The persistent behaviour observed in the FET device presented above is also manifest in the photoconductivity, as shown in the time dependent conductivity data in Figure 3. At ambient temperature the conductivity was found to increase upon illumination by about 0.06 Sm$^{-1}$ over the course of several minutes, and after shading the film its dark conductivity fell very slowly, initially following a non-exponential time dependence before approaching a simple exponential form when the excess conductivity was sufficiently small (see Fig. 3). The behaviour clearly points to the existence of electron traps that are filled under illumination, and which are slowly thermally depopulated after the light is removed.

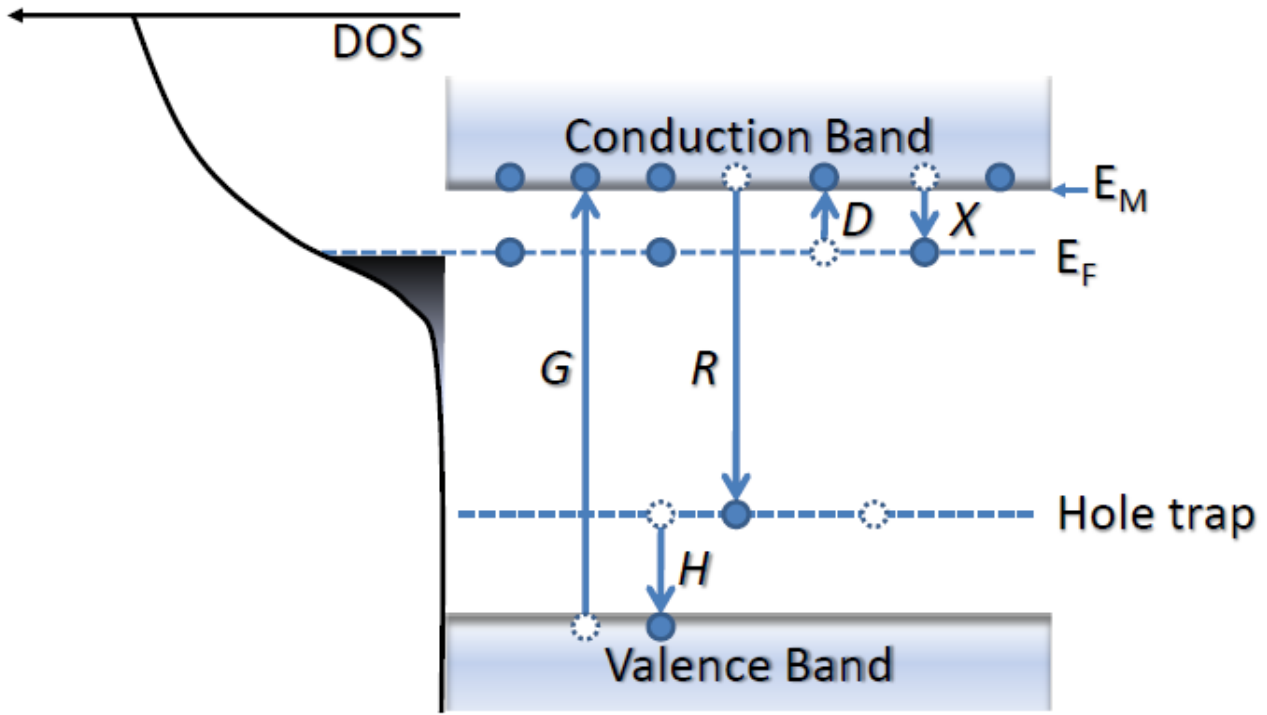

Figure 4: Processes contributing to the photoconductivity in GdN. *G*: optical excitation from the valence band to the conduction band; *H*: hole capture; *R*: recombination of an electron from the conduction band with a trapped hole; *X*: electron trapping by a nitrogen vacancy state at the Fermi level ($E_F$); *D*: electron de-trapping.

To understand the photoconductivity note first that the nitrogen vacancies form a tail of localized impurity states (the "traps") below the conduction band mobility edge ($E_m$).[28] The occupation balance between localized tail states and extended band states is then determined by $E_m$ and the Fermi level $E_F$; the activation energy measured above is $E_m$-$E_F$. The relevant processes determining the photoconductivity in the strongly *n*-type material are then illustrated in Figure 4.[29] The incident light excites electrons from the valence to conduction band (*G*), with the resulting hole very quickly captured by a deep level forming a recombination centre (*H*). Electrons in the conduction band interact with the impurity levels via trapping (*X*) and thermally excited de-trapping (*D*), which we assume takes place at a sufficient rate that thermal equilibrium is reached between the conduction band and impurity levels. Recombination (*R*) takes place between electrons in extended conduction-band states and hole traps, and especially in the limit of small concentration of empty hole traps this is the rate-limiting process. Clearly steady-state is reached when $G = R$ and $D = X$.

It is the concentration $n_c$ of the extended-state electrons in the CB that determines the conductivity; the concentration $n_t$ of trapped electrons serves only as a pool of potential CB electrons. The occupancy of both sets of states can thus be described by a quasi-Fermi level ($E_F^*$), and the photoconductivity can be interpreted in terms of a rise in $E_F^*$ as electrons are photo-excited into the conduction band. As temperature decreases the Fermi function becomes sharper, so the shift in the quasi-Fermi level required to introduce a given density of excess carriers ($\delta n_c$) introduces more excess trapped charges ($\delta n_t$) at low temperature than it does at high temperature. The density of recombination centres $\rho$ is given by $\rho = \delta n_c + \delta n_t$. Therefore, there is a larger density of recombination centres per CB electron at low temperature than at high temperature, and thus the photoconductivity decreases with temperature. For our GdN films this decrease occurs only below about 150 K. At higher temperatures the Fermi function is broad enough that the trap levels remain essentially empty, i.e., all dopant levels are activated. The photoconductivity increases as the temperature falls below $T_C$, corresponding to a decrease in the depth of the trap levels as the conduction band undergoes exchange splitting in the ferromagnetic state.[2,4,5,25-27] At the lowest temperatures there is again evidence of decreasing photoconductivity, repeating the behaviour observed just above $T_C$. Thus the data signal that the $V_N$ level remains below the conduction band in the ferromagnetic state.

The non-exponential decrease of the conductivity back towards the dark conductivity after optical excitation is also readily understood. Electrons must be thermally excited into the conduction band to recombine, but this decreases the quasi-Fermi level, and so the activation energy increases as the traps empty. However note that the data of Figure 3 show a simple exponential at long times as the dark conductivity is closely approached. The process is understood by noting that the density ($\rho$) of empty hole traps fills at the recombination rate,

$$\frac{d\rho}{dt} \propto R = -\Gamma n_c \rho, \tag{1}$$

where $\Gamma$ is a constant. Thus in the limit of $n_c$ asymptotically approaching its final dark value, $\rho$ follows a decaying exponential time dependence.

The conductivity follows $n_c$ rather than $\rho$, so it remains to demonstrate that the excess carrier density, $\delta n_c$, follows the same time dependence. As noted above the electrons from the empty hole traps appear as excess electrons in the CB and electron traps, so $\rho = \delta n_c + \delta n_t$. $\delta n_c$ and $\delta n_t$ are both determined by $E_F^*$, and as this approaches the dark equilibrium Fermi energy their first-order values are

$$\delta n_c = n_c \left( E_F^* - E_F \right)/k_B T \tag{2}$$

$$\delta n_t = \mathcal{D}(E_F)\left( E_F^* - E_F \right), \tag{3}$$

where $\mathcal{D}(E_F)$ is the density of localized states at the Fermi level. Thus at constant temperature

$$\delta n_c = \rho/\left[1 + \mathcal{D}(E_F)k_B T / n_c\right] \tag{4}$$

follows the same exponential dependence as does $\rho$.

In summary, we have demonstrated field-effect tuning of the carrier concentration in GdN, opening the possibility of investigating the role of the carrier concentration in determining the fundamental properties of GdN. The results point to the presence of electron trapping states most likely associated with nitrogen vacancy levels. The GdN films display a clear photoconductivity that allows further probing of the trapping states. There is no signature at all of a carrier-mediated contribution to inter-ion exchange.

**Acknowledgements**

We acknowledge financial support from the NZ FRST (Grant No. VICX0808) and the Marsden Fund (Grant No. 08-VUW-030). The MacDiarmid Institute is supported by the New Zealand Centres of Research Excellence Fund. We also thank Tanmay Maity for magnetic measurements which support these results and Natalie Plank for experimental assistance.